# Localized NMR Mediated by Electrical-Field-Induced Domain Wall Oscillation in Quantum-Hall-Ferromagnet Nanowire

S. Miyamoto,*,†,§ T. Miura,† S. Watanabe,‡ K. Nagase,† and Y. Hirayama*,†,¶

*Department of Physics, Tohoku University, 6-3 Aramaki Aza Aoba, Aoba-ku, Sendai 980-8578, Japan, Institute of Science and Engineering, Kanazawa University, Kanazawa 920-1192, Japan, and WPI-AIMR, Tohoku University, 2-1-1 Katahira, Aoba-ku, Sendai 980-8577, Japan*

E-mail: miyamoto.research@gmail.com; hirayama@m.tohoku.ac.jp

**ABSTRACT** We present fractional quantum Hall domain walls confined in a gate-defined wire structure. Our experiments utilize spatial oscillation of domain walls driven by radio frequency electric fields to cause nuclear magnetic resonance. The resulting spectra are discussed in terms of both large quadrupole fields created around the wire and hyperfine fields associated with the oscillating domain walls. This provides the experimental fact that the domain walls survive near the confined geometry despite of potential deformation, by which a localized magnetic resonance is allowed in electrical means.

---

*To whom correspondence should be addressed
†Tohoku University
‡Kanazawa University
¶WPI-AIMR
§Present address: School of Fundamental Science and Technology, Keio University, 3-14-1 Hiyoshi, Kohoku-ku, Yokohama 223-8522, Japan

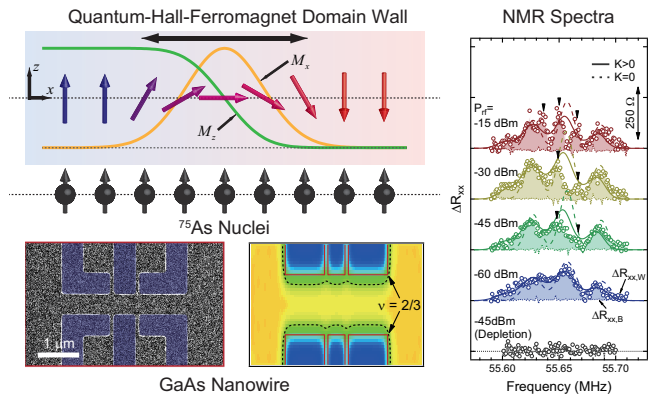

TOC Graphic For Nano Letters

Early exploration of domain-wall physics in nanoscale structures has been made mainly using ferromagnetic metals[1] and ferromagnetic semiconductors.[2] In contrast, a two-dimensional electron system (2DES) in semiconductors is also known to exhibit the state reminiscent of a ferromagnetic class under low temperature and high magnetic fields.[3–5] In particular, a fractional quantum Hall (FQH) regime represents a spin phase transition driven by the competitive interplay of the Zeeman and Coulomb energies.[6,7] The spin-related phenomenon in the FQH regime is interpreted by the single-particle picture of a composite fermion (CF), which is an electron attached to an even number of magnetic flux quanta.[8] The Zeeman energy has a linear dependence on magnetic fields, while the CF Landau levels are determined by the Coulomb energy. Due to the inverse relation of the Coulomb energy with the interparticle distance or the magnetic length, it is in principle characterized by the carrier density and/or the magnetic fields. At the electronic filling factor of $\nu = 2/3$ corresponding to $\nu_{CF} = 2$ in the CF picture,[9] both spin-split states of the lowest Landau level $(0,\uparrow)$ and $(0,\downarrow)$ are occupied at small Zeeman energy, which produces a spin-unpolarized phase of $\uparrow\downarrow$. When the Zeeman energy exceeds the Coulomb energy, the spin phase undergoes a transition to a spin-polarized phase of $\uparrow\uparrow$ by filling the spin-up states of the first and second Landau levels $(0,\uparrow)$ and $(1,\uparrow)$. As such a spin phase transition is considered of first order nature, it is proven by nuclear magnetic resonance (NMR) that these two spin phases coexist, thereby building the domain structures of electron spins.[10] At the boundary of the divided phases, namely, domain wall (DW), dynamic nuclear polarization (DNP) is accompanied via a frequent flip-flop process of electron spins.[11,12] A comprehensive study involving nuclear spins has hitherto unveiled exotic electron charge and spin states at other filling factors; for instance, the formation of Wigner crystal around $\nu = 2$ and $3$ [13,14] and stripe phase at $\nu = 5/2$,[15] the low-energy collective excitation of skyrmions around $\nu = 1$,[16–18] the electron spin polarization at $\nu = 1/2$ [19,20] and $\nu = 5/2$,[21,22] and the canted antiferromagnet state at a bilayer filling factor $\nu_{tot} = 2$.[23,24] Meanwhile, though various efforts to investigate the domain structures at $\nu = 2/3$ have been devoted from experimental and theoretical aspects,[25–28] their microscopic features and local spin properties remain indistinct points to be addressed. Recently, a novel NMR, called nuclear electric resonance (NER), was demonstrated on the basis of the spatial oscillation of DWs induced by gate electric fields, providing valuable insight into the DWs themselves.[29,30] For the sake of a nonzero in-plane component of hyperfine fields in the phase transition region, the DW oscillation gives rise to the NMR for neighboring polarized nuclei. In this Letter, by applying this NER scheme to fine-gate structures, we show that the fragile state of DWs is selectively formed in a gate-defined wire. This indicates that the nanoscale NMR is enabled by means of local gate electric fields, which can be inversely utilized to detect the confined DWs. Such a localized NER might lead to probe single DWs in the FQH regime and further explore the internal electron spin states.

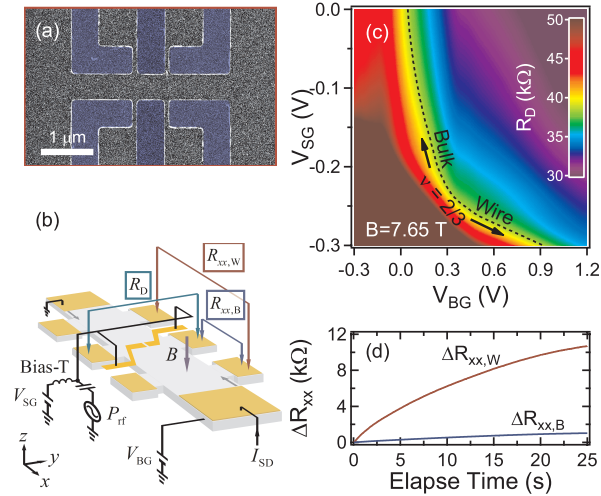

Figure 1: (a) Scanning electron micrograph of the wire with the split gates of triple pairs highlighted by a blue color. (b) Measurement configuration for the localized NER experiments. Respective resistances are monitored with a standard lock-in technique. The magnetic field points in the $-z$ direction and the flow direction of the edge current is represented by the gray arrows on the Hall bar. A rf signal superimposed onto a dc bias $V_{SG}$ by a bias tee is fed to the split gates. (c) Two-dimensional plot of $R_D$ as a function of $V_{SG}$ and $V_{BG}$. The dotted line denotes the diagonal resistance of $R_D^{(\nu=2/3)}$. (d) Time evolution of $\Delta R_{xx,W}$ and $\Delta R_{xx,B}$ at $\nu_W \approx 2/3$ [$(V_{SG}, V_{BG}) = (-0.275 \text{ V}, 0.500 \text{ V})$]. Note the horizontal axis is shown on the second time scale corresponding to a characteristic time scale of nuclear spins.

The wafer used in this study consists of a 20 nm-thick GaAs quantum well located at a position of 175 nm from the surface. A heavily doped GaAs buffer layer 1 $\mu$m below the quantum well provides a back gate, and a $\delta$-doping layer is embedded at a setback distance of 100 nm on the front side. Low-temperature Hall measurements show a mobility of $1.0 \times 10^6$ cm$^2$/V s and a carrier density of $1.2 \times 10^{15}$ m$^{-2}$ at zero back-gate voltage ($V_{BG} = 0$ V). The split gates of triple pairs fabricated on the wafer are biased at the same voltage of $V_{SG}$, thereby defining a wire structure with a lithographic size of 3 $\mu$m in length and 500 nm in width [Figure 1a]. The wire structure is designed at the center of two aligned Hall probes as illustrated in Figure 1b. The whole of sample is kept at a temperature of 100 mK in a top-loading dilution refrigerator and subject to a perpendicular magnetic field of $B = 7.65$ T. Tuning of split-gate and back-gate voltages ($V_{SG}$, $V_{BG}$) under the fixed magnetic field settles a local filling factor in the wire $\nu_W$, which is evaluated from a four-terminal diagonal resistance measured across the wire, $R_D(= h/e^2(1/\nu))$.[31–34] While for the localized NER a radio frequency (rf) signal is applied to the split gates (instead of the back gate), we concurrently record two longitudinal resistances; $R_{xx,W}$ probed across the wire region and $R_{xx,B}$ reflecting the surrounding bulk regions. By comparing between the changes in these two resistances, $\Delta R_{xx,W}$ and $\Delta R_{xx,B}$, we identify if the observed spectra are derived from the localized NER around the wire region.

Figure 1c displays $R_D$ measured at different ($V_{SG}$, $V_{BG}$). As marked by the dotted line, a contour line of the anticipated diagonal resistance, $R_D^{(\nu=2/3)} = 38.7$ k$\Omega$, is significantly curved with decreasing $V_{SG}$. Then, the place where the $\nu = 2/3$ state develops gradually shifts from the bulk regions ($\nu_B \approx 2/3$) to the wire region ($\nu_W \approx 2/3$). In order to achieve the domain structures in the wire region, ($V_{SG}$, $V_{BG}$) is set around the graph area adjacent to the curved line. At $\nu_W \approx 2/3$, applying a high ac current of $I_{SD} = 100$ nA causes a huge enhancement in $\Delta R_{xx,W}$ as shown in Figure 1d, which is in contrast with a slight increment in $\Delta R_{xx,B}$. This supports that the DNP tends to take place around a narrow constriction by virtue of a local increase in the current density.[35,36] Subsequently, the rf signal is applied to the split gates for the localized NER. When a frequency of the rf signal is swept at around the Larmor frequency of $^{75}$As nuclei, a typical spectrum is obtained as shown in Figure 2a. Remarkably, a clear signal can be observed in $\Delta R_{xx,W}$, whereas $\Delta R_{xx,B}$ hardly exhibits a change. Thus, the acquired signal evidently originates from localized oscillation of DWs around the wire region. In Figure 3a, the integrated intensity of the NER spectra observed at various ($V_{SG}$, $V_{BG}$) is put together as a circle size upon each curve of $R_D$. Obviously, the graph area where the intense signals are detected spreads near the dotted line signifying $R_D^{(\nu=2/3)}$ even though the filling factor in the bulk regions is rendered $\nu_\text{B} = 0.82$ to $1.01$. In addition, the signals can be still observed far above the dotted line. These experimental results suggest that the DWs can form in the wire as well as at the wire entrances having a gentle potential slope. The compressible DWs provide a backscattering path between counter-propagating edge channels, whereas the tunneling of quasiparticles recently investigated[32,33] needs to occur across the gapped domains. It seems that the backscattering process via the DWs governs the tunneling process across the domains. Additionally, the ac current applied for the DNP is large enough to cause the breakdown of the FQH states in a narrow constriction.[34] Therefore, the $R_\text{D}$ measurements in Figures 1c and 3a show no plateaus related to $\nu = 2/3$, probably due to the backscattering or the breakdown enhanced in the wire.

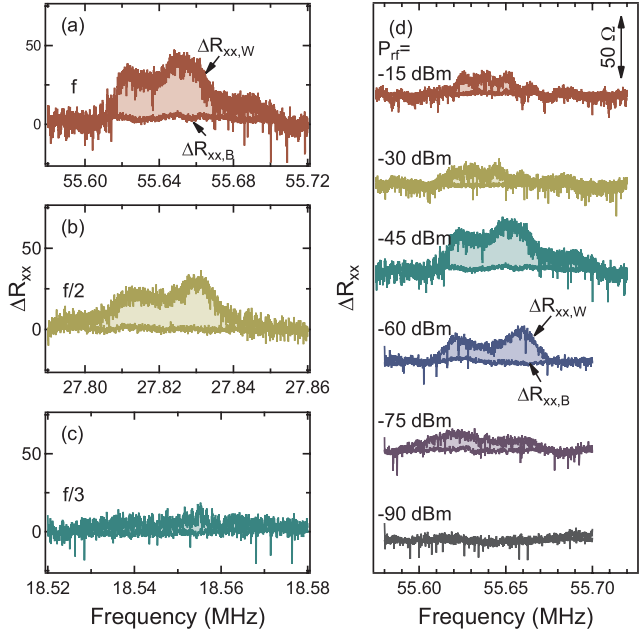

Figure 2: Localized NER spectra for $^{75}$As detected by the difference between $\Delta R_{xx,\text{W}}$ and $\Delta R_{xx,\text{B}}$. $\nu_\text{W} \approx 2/3$ is fixed by setting $(V_\text{SG}, V_\text{BG}) = (-0.250\text{ V}, 0.500\text{ V})$. When the rf power is fixed at $P_\text{rf} = -45$ dBm, (a) panel is obtained at around the Larmor frequency $f$, and the (b) and (c) panels are for the subharmonic frequencies of $f/2$ and $f/3$, respectively. (d) Power dependence of the localized NER spectra for the fundamental resonance. The traces are vertically offset for clarity.

Figure 3b represents the local filling factors around the wire that are based on the carrier density obtained by self-consistently solving the Schrödinger-Poisson equations under zero magnetic field. Hence, the numerical calculation confirms that the $\nu \approx 2/3$ state forms around the wire including the outside gate-edge region while the different state of $\nu_B \approx 0.82$ forms in the bulk regions. This result agrees with the carrier density measured in the wire and bulk regions at low magnetic fields, and then the separation between the dotted contour lines approximately yields an effective wire width of $\sim 330$ nm. However, the edge reconstruction at $\nu = 2/3$ results in the neutral modes propagating upstream apart from the charge mode propagating downstream.[37] More recently, it has been reported that near $\nu_B \approx 1$ a smooth potential at the gate-defined edge causes an incompressible stripe of $\nu = 2/3$ proximally accompanying the neutral modes.[38] Thus, the domain structures may form in the gate-edge region where the DW oscillation contributes to the localized NER signal. Indeed, the gate-edge contribution to the localized NER signal is ensured to be practically nought (see Supporting Information S1). The NER experiment at $\nu_B \approx 2/3$ contains the gate-edge contribution, which results in the signal appearance in both $\Delta R_{xx,B}$ and $\Delta R_{xx,W}$. By contrast, the signature for the localized NER manifests itself only in $\Delta R_{xx,W}$ at $\nu_W \approx 2/3$.

When the DW moves back and forth in the lattice, the internal electron spins cause a rotating hyperfine field. Then the fixed nuclei experience the pulse-like and irregular fields generated by the in-plane hyperfine component.[29] The spectral density of the effective fields contains higher-order components so that the observability at subharmonic resonance frequencies is a unique characteristic of the NER.[30] As shown in Figures 2b,c, the localized NER can be similarly observed at the subharmonic frequencies apart from the fundamental resonance. As the higher-order components of the spectral density are smaller, the detected intensity is reasonably lowered at the subharmonic resonances. With regard to an out-of-plane hyperfine component, as discussed in refs 29 and 30, the nuclei feel the time-averaged and blurred fields created by the two distinct domains when the DWs are oscillated with large amplitude. In Figure 2d, the spectral width becomes gradually reduced with increasing the rf power $P_{rf}$. The signal can be detected at a small rf power of $P_{rf} = -75$ dBm, but the signal intensity unexpectedly drops above $P_{rf} = -30$ dBm. The latter implies that the DWs, which are necessary in both the processes of the DNP and NER, cannot stably form at such high rf powers. More specifically, because the current is considered to pass along the DWs[39,40] and polarize the nearby nuclear spins,[11,12] a continuous motion of DWs interrupts the DNP for the fixed nuclei. In addition, under the application of the large current for the DNP, the partial breakdown of the domain structures possibly makes the NER inefficient.

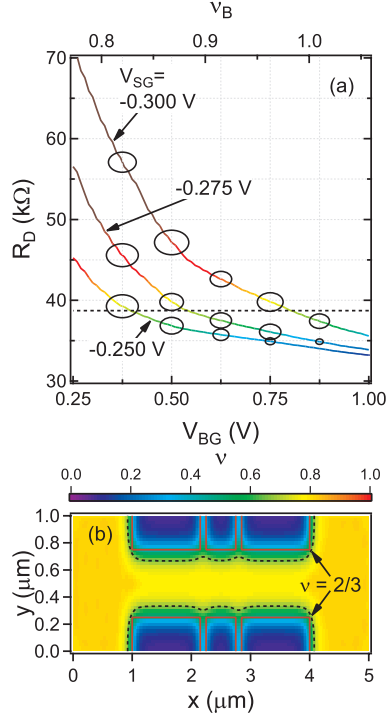

Figure 3: (a) Integrated intensity plot of the localized NER spectra for $^{75}$As along the $R_D$ curves obtained by sweeping $V_{BG}$ at three selected values of $V_{SG}$. The top axis is given by $\nu_B$ and the dotted line denotes the corresponding level to $R_D^{(\nu=2/3)}$. No discernible signal can be detected in the change in $\Delta R_{xx,B}$ at any investigated combination of ($V_{SG}$, $V_{BG}$). Regarding $\Delta R_{xx,B}$ as a reference signal, the integrated intensity is here yielded by the spectral area enclosed by $\Delta R_{xx,W}$ and $\Delta R_{xx,B}$. (b) Calculated filling factor in the vicinity of the wire at a representative voltage set of ($V_{SG}$, $V_{BG}$) = ($-0.250$ V, $0.375$ V). The dotted lines indicate the contour line of $\nu = 2/3$, and the red lines show the outline of the triple-paired split gates.

In order to deconvolute the DNP and NER into separate steps, we carry out pump-probe measurements (see Figure 4a). First, the DNP is started by injecting a high current of $I_{SD} = 100$ nA for 100 s. Immediately after that, $R_{xx,W}$ and $R_{xx,B}$ are measured once. The NER is subsequently invoked by activating the rf signal for 5 s, while the current is switched off.[41] Then, the both resistances are measured again and the current in the two read steps is lowered to $I_{SD} = 10$ nA. Here, $\Delta R_{xx,W}$ and $\Delta R_{xx,B}$ are redefined as an amount of the resistive change before and after the NER. This sequence is repeated for different frequencies to obtain a NER spectrum. Figure 4b plots the rf power dependence of the NER spectrum taken in this manner. Similarly to the prior continuous-wave experiments, $\Delta R_{xx,\text{B}}$ shows no variation. On the contrary, the intense signal of $\Delta R_{xx,\text{W}}$ can be maintained even above $P_{\text{rf}} = -30$ dBm, and the power-dependent spectra observed here intriguingly comprise several peaks. Such rich features cannot be sufficiently resolved due to the signal decay in Figure 2d. Besides, during the rf irradiation, we apply a negatively large $V_{\text{SG}}$ enough to deplete electrons around the wire region. Then the signal of $\Delta R_{xx,\text{W}}$ is found to completely disappear as shown by the lowermost part in Figure 4b. Since the DWs are nonexistent under the depletion circumstance, the signal vanishing presents the signature of the localized NER. This is the major reason why our observation is distinguished from the conventional NMR, which does not rely on the DW occurrence.

Fine-gate structures such as the wire distort the lattice close to a surface, locally creating the quadrupole fields for the nuclei in the QW. In parallel, the out-of-plane component of the hyperfine fields is seen as an effective magnetic field along the direction of the external field, leading to the Knight shift for the nuclear spin states. Taking into account both contributions of the quadrupole fields and the hyperfine fields, the nuclear spin Hamiltonian is described as $H = H_Z + H_Q + H_K$ by using the Zeeman term of $H_Z = -\hbar\omega_Z I_z$. $H_Q = \frac{\delta}{6}\left[3I_z^2 - I(I+1)\right]$ is the quadrupolar term and $H_K = KI_z$ is the Knight shift term. A possible energy level diagram for $^{75}$As having total spin $I = 3/2$ is illustrated as shown in Figure 4c. Under external magnetic fields, four energy states of $|m_I\rangle = |3/2\rangle, |1/2\rangle, |-1/2\rangle, |-3/2\rangle$ are equally spaced by the Zeeman energy $\hbar\omega_Z$. In the absence of the Knight shift, the quadrupole fields displace the Zeeman-split levels by $\pm\delta/2$. In addition, the spin-unpolarized domain ↑↓ makes little energy difference, whereas the spin-polarized domain ↑↑ induces the Knight shift for each quadrupole-split level. According to the NMR selection rules $\Delta m_I = \pm 1$, the following spin transitions are allowed as a consequence; $\hbar\omega_Z - \delta$, $\hbar\omega_Z - \delta - K$, $\hbar\omega_Z$, $\hbar\omega_Z - K$, $\hbar\omega_Z + \delta$, $\hbar\omega_Z + \delta - K$. This indicates that the localized NER spectra could be split into six peaks unless $K$ takes a multiple of $\delta$. In practice, since the Knight shift is a continuously varying function of the position due to the spatial inhomogeneity, there are numerous levels in between the two states associated with the spin domains ↑↓ and ↑↑, resulting in the broadening of the spectral peaks. The absorption spectrum for an individual nucleus is here regarded as the

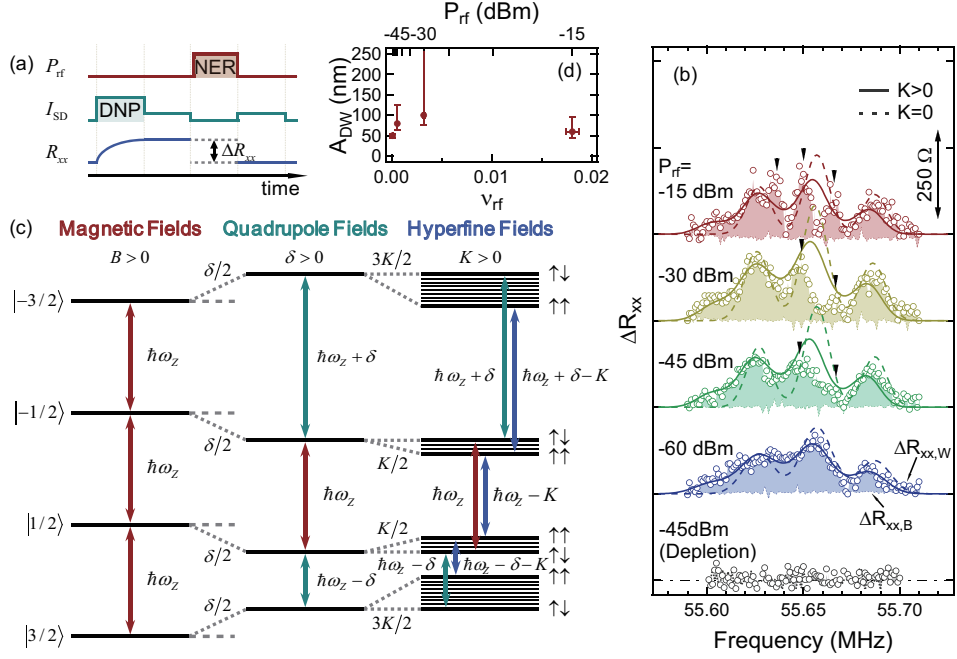

Figure 4: (a) Sequence of the pump-probe measurements used for the localized NER. (b) Localized NER spectra for $^{75}$As at different $P_{rf}$. $\nu_W \approx 2/3$ is fixed by setting $(V_{SG}, V_{BG}) = (-0.250$ V, $0.500$ V). The open circles are obtained by $\Delta R_{xx,W}$, and the dotted spectral baselines are given by $\Delta R_{xx,B}$. The solid lines are the simulated curves based on the model in the text using $f_0 = 55.657$ MHz and $\Gamma = 9.0$ to $11.5$ kHz. The dashed lines correspond to the case free from the Knight shift, $K = 0$. For reference, the plots taken under the depletion condition ($V_{SG} = -0.6$ V) are shown in the lowermost part. The traces are vertically offset for clarity. (c) Schematic energy level diagram of nuclear spin states for $I = 3/2$ which in turn incorporates the external magnetic fields (left), the quadrupole fields around the split-gate structure (middle), and the hyperfine fields created by the spin domains of ↑↑ and ↑↓ (right). (d) DW oscillation amplitude as a function of a filling factor magnitude effectively modulated by the rf signal. The top axis shows the corresponding $P_{rf}$.

Gaussian function $g(f;x,z) = \exp\left\{-[f - f_0 - m\delta/h + K(x,z)/h]^2/\Gamma^2\right\}$ with the nuclear dipolar broadening of $\Gamma$, where $f_0 = \omega_Z/2\pi$ and $m = -1, 0, +1$. Then the spectral line shape can be given by calculating the Knight shift at each nuclear location and integrating the above function over all nuclei in the wire (see Supporting Information S2).

The calculated NER spectra are shown by the solid lines in Figure 4b. At the lowest rf power of $P_{rf} = -60$ dBm, the simulated curve can reproduce the overall feature of the experimental data. Approximately 3-fold peaks arise from the quadrupole splitting, reflecting the large lattice strain accumulated around the wire. When comparing with the particular case of no Knight shift [the dashed lines in Figure 4b], it is noticed that for each quadrupole peak the Knight shift results in a spectral shoulder on a low-frequency side. The Knight shift reaches $K_{max}/h \approx 31$ kHz at a maximum, which is comparable to the quadrupole splitting of $\delta/h \approx 30$ kHz.[42] As a result the peaks overlap with each other rather than split into six. However, at the higher rf powers, the center peak begins to collapse into a couple of peaks marked by the inverted triangles [Figure 4b], gradually deviating from the calculation. The spectral peaks for intermediate frequencies cannot be thus explained solely by the quadrupole splitting and the Knight shift. In general, the transitions mediated by multiple photons ($\Delta m_I = \pm 2, \pm 3, ...$) could be allowed in a high rf power regime.[35] Two-photon processes between $|-3/2\rangle$ and $|1/2\rangle$ and between $|-1/2\rangle$ and $|3/2\rangle$ would occur in the energy between $\hbar \omega_Z \pm \delta/2$ and $\hbar \omega_Z \pm \delta/2 - K$, which manifests narrower replicas in the centers between the three quadrupole-split peaks. Furthermore, a three-photon process between $|-3/2\rangle$ and $|3/2\rangle$ would lead to a much sharper replica at the frequencies of $f_0$ and $f_0 - K/h$ regardless of the quadrupole splitting. These provide no account for the deviation from the model at the higher rf powers. Meanwhile, if considering the case of the second-order quadrupole interaction, the relevant effect can be observed mainly on the inner transition between $|-1/2\rangle$ and $|1/2\rangle$.[43] Though the additionally observed peaks near the transition might reflect strain inhomogeneity around the gate structures, further measurements are required for discussion of its origin.

On another front, $P_{rf}$ can be rephrased as an effectively rf-modulated filling factor $\nu_{rf}$ (see Supporting Information S3). In Figure 4b, it is worth noting that the localized NER even at $P_{rf} = -60$ dBm brings about a strong signal from the wire region, which is mediated by a slight modulation of $\nu_{rf} \approx 10^{-4}$. In the case of the back-gate NER using the 50 $\mu$m-width Hall bar, the bulk signal is difficult to detect below $P_{rf} = -30$ dBm, i.e. $\nu_{rf} \approx 10^{-3}$. Hence the higher-sensitivity measurement in a local region is assisted probably by the constriction of the wire structure. As a function of $\nu_{rf}$, Figure 4d plots the DW oscillation amplitude $A_{DW}$ that is employed as a calculation parameter for reproducing the experimental data. While $\nu_{rf}$ exponentially grows with increasing $P_{rf}$, $A_{DW}$ seems to remain limited below a few hundreds of nm at $\nu_{rf}/(2/3) \approx 1\%$. Recently, the real-space imaging of the domain structures showed that a 0.3% change in the filling factor displaces the DWs in a micrometer scale, thereby triggering the spin phase transition for negligible nuclear polarization.[26]

In general, when the DNP is induced by current injection, the resistance peak associated with the spin phase transition is broadened with respect to the filling factor.[10–12] Since a larger change in the filling factor is then required for the spin phase transition, it is possible that the DWs become insensitive to the filling factor at a high degree of nuclear polarization. In this case we can infer that the above difference in the DW mobility could be related to the polarization degree of nuclear spins close to the moved DWs.

The DW oscillation has previously been used for the first observation of the NMR in a ferromagnetic material.[44] Although in the earliest work the DW displacement results from a magnetic-field modulation, the electric-field modulation causes the equivalent effect via the filling factor. Hence the NER has a close relation to the NMR in the ferromagnetic system. Whereas many exciting experiments of the controlled DWs have been reported in nanoscale magnetic devices as stated in the beginning, the studies on the QH ferromagnetism are mostly limited to investigation of its bulk properties. Recently, in order to control the FQH states with exotic properties, efforts to study them in a small region have been made.[32,33] Although to our knowledge no experiments have been reported in which the DWs take a Bloch type or a Neel type at $\nu = 2/3$, the DW switching between the two types might be triggered in the confinement similar to in a ferromagnetic nanowire.[45] In these contexts, our experiments provide NMR evidence that a fragile state of DWs can survive in the confinement, which offers a key result to advance the QH ferromagnetism from a bulk regime to a nanoscale regime. However, it is well-known that nanoscale NMR experiments based on optical detection have been intensively conducted using nitrogen-vacancy (NV) centers in diamond.[46] While such atomic defects achieve extremely high resolution in probed space, the hyperfine interaction may enhance the locality for the NER in this experiment. Even though the NV centers can operate as an ultrasensitive magnetic sensor to study domain structures in a surface thin film, they are still challenging to use for detecting the QH domains that occur in a deeply embedded quantum well under limited condition. Furthermore, although antenna circuits are routinely necessary to supply strong magnetic fields for a fast spin operation, the electrical gates defining nanostructures can be utilized through mediated application of hyperfine fields.

In conclusion, the combination of the DNP and NER has been demonstrated in a local region of the wire. While the basic behavior is similar to that which occurs in the bulk NER, the localized NER signal is found to be significantly enhanced. In addition, we showed that the signal spectra, distinct from the bulk NER spectra, are analyzed by taking into account the hyperfine fields associated with the oscillating DWs as well as the quadrupole fields around the wire. Although the additional splitting is observed at the higher rf powers, the large quadrupole splitting has advantage to control nuclear spin systems possessing $I > 1/2$. From the Knight shift for $^{75}$As, the out-of-plane hyperfine fields are estimated to be up to 4.2 mT. Under the assumption of an isotropic hyperfine interaction, it is conceivable that in the in-plane direction a comparable magnitude of rf-modulated fields is irradiated only onto the nuclei exposed by the DW oscillation. Hence, the all-electrical NMR scheme based on the electrical-field induction and resistive detection may be useful for a sensitive probe of single DWs and provides an efficient principle for local gate manipulation of nuclear spins.

**Supporting Information Available**

Additional descriptions for the gate-edge contribution to NER signal, the model calculation of NER spectral line shape, and the effective modulation of local filling factor.

**Author Contributions**

S.M., T.M., and K.N. designed and fabricated the nanowire device. S.M., T.M., and S.W. performed the measurements and the analysis. The manuscript was written through contributions of all authors.

**Notes**

The authors declare no competing financial interest.

## Acknowledgement

The high-quality wafers used for this work were supplied by K. Muraki at NTT Basic Research Laboratories. The authors are grateful to K. Akiba, T. Hatano, G. Yusa, J. N. Moore, T. Tomimatsu, and K. Hashimoto for valuable discussions. The reported experiments were carried out in ERATO Nuclear Spin Project supported by JST. Y.H. acknowledges support from MEXT KAKENHI Grant Numbers 15H05867 and 26287059.

cal value reported in the bulk ($\sim$ 7 kHz)[24] and close to a huge value in a depleted constriction device ($\sim$ 54 kHz).[35] The similar estimations assuming the quadrupole fields yield these values, which can be compared with the present work.

# Supporting Information:

# Localized NMR Mediated by Electrical-Field-Induced

# Domain Wall Oscillation in

# Quantum-Hall-Ferromagnet Nanowire

S. Miyamoto,[*,†,§] T. Miura,[†] S. Watanabe,[‡] K. Nagase,[†] and Y. Hirayama[*,†,¶]

*Department of Physics, Tohoku University, 6-3 Aramaki Aza Aoba, Aoba-ku, Sendai 980-8578, Japan, Institute of Science and Engineering, Kanazawa University, Kanazawa 920-1192, Japan, and WPI-AIMR, Tohoku University, 2-1-1 Katahira, Aoba-ku, Sendai 980-8577, Japan*

E-mail: miyamoto.research@gmail.com; hirayama@m.tohoku.ac.jp

## S1: Gate-Edge Contribution to NER Signal

The additional NER experiments are performed by rendering the filling factor in the bulk (not in the wire) $\nu_B \approx 2/3$. When the rf signal is added to the back gate (BG-NER) or the split gates (SG-NER), the respective NERs exhibit the spectra in Figure S1. The BG-NER is mediated by the oscillation of numerous DWs in the bulk, showing the double-peak spectra derived from the hyperfine fields of the two spin phases. There is no need to consider the quadrupole splitting because the nuclear spins feeling the local strain fields around the split-gate structures hardly contribute to the signal collected from the whole bulk region. On the other hand, since the SG-NER results from the potential edge along the split-gate structures, 3-fold peaks appear due to the quadrupole splitting. In addition, the SG-NER experiment suggests that when the DW oscillation at potential edge becomes involved in the NER, the signals should be detectable in both $\Delta R_{xx,\text{B}}$ and $\Delta R_{xx,\text{W}}$. However, as addressed in the text, the localized NER around the wire region shows no clear variation in $\Delta R_{xx,\text{B}}$. Hence the edge state of $\nu = 2/3$ outside the wire makes little contribution to the localized NER signal in the present case.

---

[*]To whom correspondence should be addressed
[†]Tohoku University
[‡]Kanazawa University
[¶]WPI-AIMR
[§]Present address: School of Fundamental Science and Technology, Keio University, 3-14-1 Hiyoshi, Kohoku-ku, Yokohama 223-8522, Japan

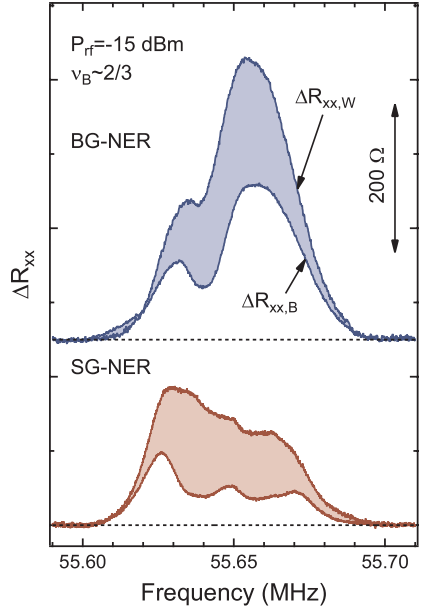

Figure S1: BG-NER and SG-NER spectra for $^{75}$As obtained at $\nu_\text{B} \approx 2/3$ while continuously applying the rf singal of $P_\text{rf} = -15$ dBm.

## S2: Model Calculation of NER Spectral Lineshape

We introduce an analytic description of the DW profile in a one-dimensional direction. Given that the in-plane magnetization in the DW is a simple Gaussian function, $M_x(x) = \exp\left[-4\ln(2)(x-x_0)^2/W^2\right]$, the out-of-plane magnetization is determined by $M_z(x) = -\left[\text{sgn}(x-x_0)\sqrt{1-M_x^2} - 1\right]/2$. As shown in Figure S2a, a DW having a width of $W$ and center position of $x_0$ is oscillated with an amplitude of $A_\text{DW}$ in a sinusoidal fashion. Then, the Fourier transform of the time-varying $M_x$ offers the spectral density specified at each in-plane position $S(x)$ for the fundamental resonance [Figure S2b]. On the other hand, the nuclear spin polarization generated by the DNP process around the DWs diffuses away from the DWs into the domains. Recently, we found that the nuclear spin polarization spreads over a wide area and results in a relatively homogeneous distribution inside the domains.[1] In this regard, the nuclei inside the DW are most likely to be relaxed due to the low-frequency spin fluctuations inherent in the DW.[2–4] This leads us to approximate the distribution of nuclear polarization by $|P_N(x)| = P_0(= \text{const.})$ for $|x| > W/2$ and $|P_N(x)| = 0$ for $|x| \leq W/2$. Based on the methods described in refs 5-8, the NER spectral intensity is expressed in the integral form over all nuclei exposed to the DW oscillation which are contained in the quantum well of width $w$;

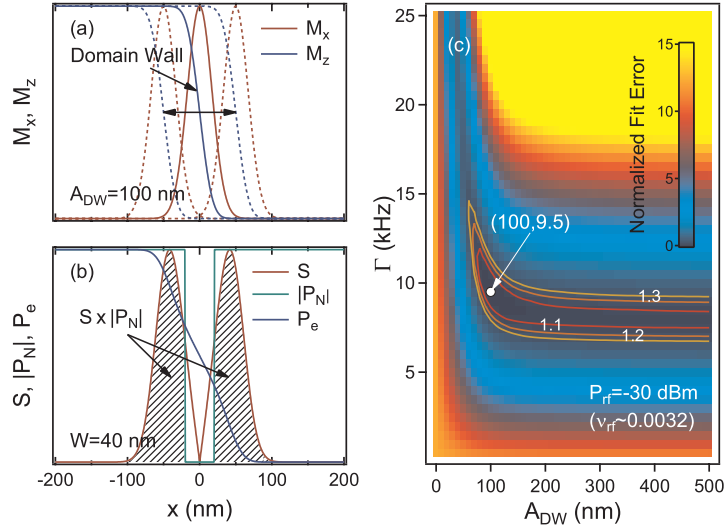

Figure S2: (a) In-plane and out-of-plane magnetization components, $M_x$ and $M_z$, in the phase transition region near the DW. The dotted lines show the two components at odd quarter oscillation cycles. (b) Spectral density $S$, nuclear spin polarization $|P_N|$, and electron spin polarization $P_e$ around the oscillating DW. The shadow area of $S \times |P_N|$ denotes the distribution of the nuclei depolarized by the DW oscillation. (c) Map of the normalized least-square error between the model calculation and the experimental data.

$$I(f) \propto \int_{-\infty}^{\infty} dx \int_{-w/2}^{w/2} dz \sum_{m=-1}^{1} A_m g(f;x,z) S(x) |P_N(x)|. \tag{S1}$$

Then an individual spin packet is regarded as the Gaussian function of

$$g(f;x,z) = \exp\left\{-[f - f_0 - m\delta/h + K(x,z)/h]^2/\Gamma^2\right\} \tag{S2}$$

where $f_0 (= \omega_Z/2\pi)$ is the bare unshifted resonance frequency, $\delta$ the quadrupole splitting, and $\Gamma$ the nuclear dipolar broadening, respectively. The Knight shift is reduced to $K(x,z)/h = \alpha_{\text{full}} P_e(x) |\psi_e(z)|^2 \beta(x)$ with $\alpha_{\text{full}}$ representing the Knight shift size expected for full polarization. $P_e(x)$ is the electron spin polarization across the oscillating DW that is obtained by time-averaging $M_z$ [Figure S2b]. Additionally we consider the fact that the Knight shift also depends on the local electron density. $|\psi_e(z)|^2$ is the probability density of the wave function in the quantum well, which is numerically calculated at $V_{\text{BG}} = 0.5$ V. The spatial fluctuation in the local electron density with respect to the DW center[9,10] is represented by the last correction part of $\beta(x)$ in $K(x,z)/h$.

In the calculation, the corresponding value of $W \sim 40$ nm is used at $B = 7.65$ T since the DW width is estimated to be about 4 times as large as the magnetic length.[10] The value of $\alpha_{\text{full}} (= c \cdot n_{e0})$ is determined by the hyperfine coupling constant $c$ and the electron density $n_{e0}$. Although the experimental error depends on an uncertainty of $\pm 5\%$ in $w$, it is estimated to be about 0.32 MHz·nm for the $\nu = 2/3$ state at 6.4 T,[6] which is translated into $\alpha_{\text{full}} \sim 0.38$ MHz·nm at $B = 7.65$ T used in the present work. Meanwhile, it is shown that a tiny change of 0.3% in the in-plane electron density triggers the spin phase transition.[9] We ensured that the change in $\beta(x) (= 1 + \Delta n_e(x)/n_{e0})$ on the order of 1% has no impact on the calculated spectral lineshape. Therefore, the local variation in the electron density $\Delta n_e(x)$ is so small across the DW that its effect is negligible for the Knight shift in our measurements; $\beta(x) \simeq 1$. Moreover, the set of the relative amplitude, $(A_{-1}, A_0, A_1) = (0.6, 1.0, 0.5)$, is here chosen as unified values for the best reproduction. In order to evaluate the remaining parameters in the above model, the least-square error is calculated between the model calculation and the experimental data. For simplicity, the frequency range where the calculation largely deviates from the data, from 55.632 to 55.667 MHz, is excluded for $P_{\text{rf}} = -45, -30$ and $-15$ dBm. Figure S2c shows the least-square error mapped by taking $A_{\text{DW}}$ and $\Gamma$ as variables. The error is normalized by the minimum value which is represented at $(A_{\text{DW}}, \Gamma) = (100$ nm, $9.5$ kHz) for $P_{\text{rf}} = -30$ dBm. The values obtained in this way are applied for reproducing the experimental data in Figure 4b. The region enclosed by the contour line of 1.1

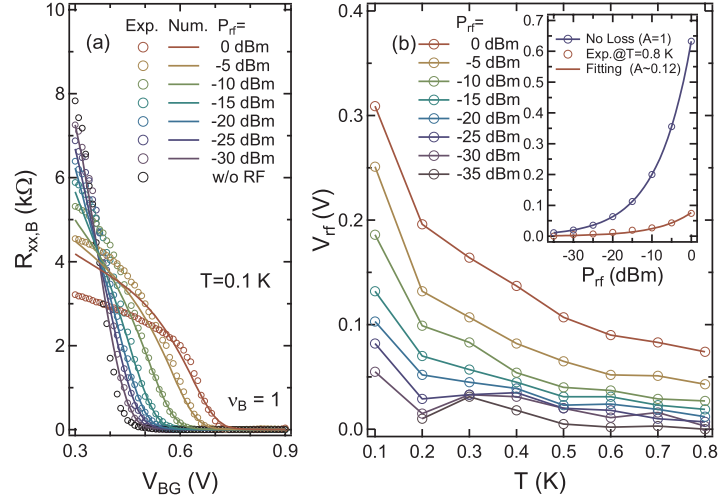

Figure S3: (a) Back-gate voltage sweep around $\nu_B = 1$ perturbed by the rf modulation at various $P_{rf}$. The circle plots show the experimental data taken at the temperature of 0.1 K and the solid lines are given by the numerical modulation. (b) Temperature dependence of the extracted $V_{rf}$ at different $P_{rf}$. The inset shows the development of $V_{rf}$ as a function of $P_{rf}$. The solid lines are given by the voltage-power relation of the rf signal terminated in an infinite impedance. The voltage attenuation is $A \sim 0.12$ in our measurement system.

on the map yields the error bars with respect to $A_{DW}$ in Figure 4d.

## S3: Effective Modulation of Local Filling Factor

The signal power launched from a rf generator, $P_{rf}$, is generally attenuated before reaching a sample in a dilution refrigerator. For a rough estimation of the effective rf voltage at the gate terminals, $V_{BG}$ sweeps of $R_{xx,B}$ around $\nu_B = 1$, characterized by a zero-resistance dip, are performed with a non-resonant rf voltage imposed on the back gate. These experimental plots obtained with varying $P_{rf}$ are compared with the reference curves given by numerically modulating and time-averaging the non-modulated sweep of $R_{xx,B}$ [see Figure S3a]. Since the experimental results include the heating effect pronounced at a higher rf power, the temperature dependence is further investigated. Consequently, a series of modulation voltage values, $V_{rf}$, used for the numerical comparisons are plotted in Figure S3b. The effective rf voltage can be extracted as $V_{rf}$ when the behavior toward a lattice temperature becomes steady. Based on the assumption that the applied $V_{rf}$ is independent of the gate geometry in the investigated frequency regime, the capacitance ratio between the back and split gates, $C_{\text{SG}}/C_{\text{BG}} \sim 3.5$, is used to eventually convert into the local filling factor modulated through the split gates.